\documentclass[twocolumn,a4paper,superscriptaddress,aps,
showpacs,nofootinbib]{revtex4}                                  
\usepackage{mathtext}
\usepackage{mathrsfs}
\usepackage{bm}
\usepackage{graphicx} 

\begin{document}

\title{Matter wave pulses characteristics}
\author{A. del Campo\footnote{E-mail: adolfo.delcampo@gmail.com}}
\author{J. G. Muga\footnote{E-mail: jg.muga@ehu.es}}
\affiliation{Departamento de Qu\'\i mica-F\'\i sica,
Universidad del Pa\'\i s Vasco, Apdo. 644, Bilbao, Spain} 

\def\la{\langle}
\def\ra{\rangle}
\def\om{\omega}
\def\Om{\Omega}
\def\vep{\varepsilon}
\def\wh{\widehat}
\def\tr{\rm{Tr}}
\def\da{\dagger}
\def\iz{\left}
\def\zi{\right}
\newcommand{\beq}{\begin{equation}}
\newcommand{\eeq}{\end{equation}}
\newcommand{\beqa}{\begin{eqnarray}}
\newcommand{\eeqa}{\end{eqnarray}}
\newcommand{\intf}{\int_{-\infty}^\infty}
\newcommand{\into}{\int_0^\infty}

\begin{abstract}
We study the properties of quantum single-particle wave pulses 
created by sharp-edged or apodized shutters with single or periodic openings.  
In particular, we examine the visibility of diffraction fringes 
depending on evolution time and temperature;  the purity 
of the state depending on the opening-time window; the accuracy of a
simplified description 
which uses ``source'' boundary conditions instead of solving an initial value
problem; and the effects of apodization on the energy width.    
\end{abstract}


\maketitle
\section{Introduction}
Diffraction in time was discussed first by Moshinsky \cite{Moshinsky52}. 
The hallmark of this phenomenon consists in temporal  
oscillations, deviating from the classical regime,
of Schr\"odinger, matter waves  
released in one or several pulses from a preparation
region in which they 
are initially confined. The original setting consisted of a sudden opening 
of a shutter to release a semi-infinite beam,
and provided a quantum, temporal analogue 
of spatial Fresnel-diffraction by a sharp edge \cite{Moshinsky52};
later more 
compicated shutter windows, initial confinements, 
and time-slit combinations have been considered \cite{Kleber}, 
in particular two temporal slits
and the corresponding interferences.  
Experimental observations have been carried out with ultracold neutrons 
\cite{HFGGGW98} and with ultracold atoms \cite{SSDD95,ASDS96,SGAD96}.
Recently, it has also been observed for electrons in a double temporal slit
experiment \cite{Lindner05}.  
The ``shutter problem'' has indeed important applications,
as it is the modelization 
of turning on and off a beam of atoms as done e.g. in
integrated atom-optical circuits 
or a planar atom waveguide \cite{SHAPM03}, and may allow us to translate 
the principles of spatial diffractive light optics to the time domain 
for matter waves \cite{BAKSZ98}.  
Besides, it provides a time-energy uncertainty relation 
\cite{Moshinsky76,Busch02} which has been verified experimentally
for atomic waves,
by realizing a Young interferometer with temporal
slits  \cite{SSDD95,ASDS96,SGAD96}. 
The Moshinsky shutter has also been discussed with the Wigner function
and tomographic probabilities \cite{MMS99}, for relativistic 
equations \cite{Moshinsky52,GRV99,DMRGV}, with dissipation 
\cite{MS01,DMR04}, in relation to Feynman paths \cite{GVY03},
or for time dependent 
barriers \cite{SK88}, including the 
time dependence of states initially confined in a box 
when the walls are suddenly 
removed \cite{GK76,Godoy02}.
With adequate interaction potentials 
added to the model,
it has been used to study
and characterize transient 
dynamics of tunnelling 
matter waves \cite{Stevens,TKF87,JJ89,BM96,GR97,GV01,GVDM02,DMRGV},
and the transient response 
to abrupt changes of the interaction potential 
in semiconductor structures and quantum dots \cite{DCM02,AP}.

Within the source boundary approximation, in which the form of the wave 
is imposed 
for all times at a source point or surface, 
many works have been carried out 
within the field of neutron interferometry \cite{GG84,FGG88}, 
considering also a triangular aperture functions \cite{FMGG90}, atom-wave
diffraction \cite{BZ97}, tunnelling dynamics \cite{Stevens,
Ranfa90,Ranfa91,Mor92,BT98,MB00,DMRGV},
and absorbing media,
in which an ultrafast peak-propagation phenomenon has been 
recently described \cite{DMR04}. 

The importance of pulse formation is nowadays enhanced due to the 
possibilities to control the aperture function of optical-shutters
in atom optics, 
and to the development 
of atom lasers. For such devices some of the first mechanisms proposed explicitly 
implied periodically switching off and on the cavity mirrors,
that is, the confining 
potential of the lasing mode. As an outcome, a pulsed atom laser 
is obtained.
Much effort has been devoted to design a continuous atom laser, 
whose principle has been demonstrated using Raman transitions \cite{MHS97,H99}. 
With this output coupling mechanism, an atom initially 
trapped suffers a transition to a nontrapped state,
receiving a momentum
kick during the process due to the photon emission. 
These transitions can be mapped  to the pulse formation, so that 
the ``continuous'' nature of the laser arises as a consequence of the
overlap of 
such  pulses.
There is in summary a strong motivation for
a thorough understanding of matter-wave 
pulse creation, even at an elementary single-particle level.

In this paper we aim to describe the characteristics of 
one-dimensional, matter-wave pulses
within 
the approximation in which interatomic interactions can be neglected, 
as it is usually 
the case in standard low-pressure atomic beams. This will be  
useful as a reference and first 
step to consider the interacting case later on. 
While the most idealized pulses have been studied  
in several works since the seminal paper of Moshinsky \cite{Moshinsky52},
some aspects of realistic pulse production have not been examined 
with enough detail yet, 
such as the effects of statistical mixing in the 
preparation beam, 
or the fact that 
optical shutters do not have absolutely sharp edges, and could  
be taylored for designing the pulse features. 
Other main objective of the paper is to compare  
two ways to model boundary conditions: as a standard initial value 
problem where the wave is specified in the preparation region at time $t=0$, 
or according to the ``source approach'' in which the wave function is
given for all times 
at the source point.

The organization of the papers is as follows:  In Sec. \ref{Mo} we review the 
Moshinsky shutter, whose main feature -diffraction in time- is quantified  
by means of the fringe visibility in Sec. \ref{vis}; 
Section \ref{evol} deals with the 
evolution of finite pulses;  
In Sec. \ref{compa} the exact results are compared with the usual ``source'' 
approximation, and, finally, in Sec. \ref{apo} the aperture function is 
modified for different
types of apodization and the  
time-energy uncertainty product is evaluated. We also consider periodical
shutter apertures and find analytical solutions for the wavefunction.

\section{Moshinsky shutter\label{Mo}}
In this section we review the ``Moshinsky shutter''
\cite{Moshinsky52,Moshinsky76} and discuss its physical interpretation.  
Consider a plane wave impinging 
from the left on a totally
absorbing infinite potential barrier located at the origin (shutter) which is 
suddenly turned off at time zero. For $t=0$ one has the initial condition
\begin{displaymath}
\psi(x,t=0)=e^{i p x/\hbar}\Theta(-x),
\end{displaymath}
where $\Theta$ is the step function, 
whereas for $t>0$ the wavefunction can be written using the free-particle 
Green's function, 
\beqa
\psi(x,t)&=&\int_{-\infty}^{0}dx'{\cal{G}}_{0}(x,t;x',0)\psi(x,0)
\nonumber
\\
&=&\frac{e^{\frac{i m x^{2}}{2 t\hbar}}}{2}\, w[-u(p,t)],  
\label{mos}
\eeqa
where 
\begin{displaymath}
{\cal{G}}_{0}(x,t;x',0)
:=\sqrt{\frac{m}{2\pi i\hbar t}}\,e^{i\frac{m(x-x')^{2}}{2\hbar t}},
\end{displaymath}
\beq\label{upt}
u(p,t):=\frac{1+i}{2}\sqrt{\frac{t}{m\hbar}}\, \left(p-\frac{m x}{t}\right),
\eeq
and
\begin{displaymath}
w(z):= e^{-z^{2}}{\rm{erfc}}(-i z)=\frac{1}{i\pi}
\int_{\Gamma_{-}}du\frac{e^{-u^{2}}}{u-z},
\end{displaymath}
is the ``$w$'' or ``Faddeyeva'' function \cite{Faddeyeva,Abram}. 
The contour $\Gamma_{-}$ goes from $-\infty$ to $\infty$ passing below the pole.

\begin{figure}
\includegraphics[height=4cm]{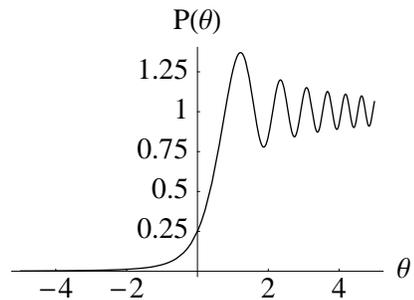}
\caption{\label{prt} 
Diffraction in time seen in $P(\theta)$, Eq. (\ref{eq:Cor}), for an
initial cut-off plane wave state. 
}
\end{figure}

\begin{figure}
\includegraphics[height=4cm]{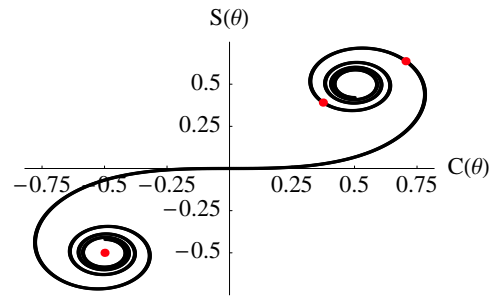}
\caption{\label{Cornu} 
Cornu espiral. The points associated with the maximum and minimum of the
largest fringe are plotted. They correspond to a maximum and a minimum of the 
distance to $(-1/2,-1/2)$. 
}
\end{figure}

The associated  
``density''\footnote{Actually $|\psi(x,t)|^2$ is a relative density  
with respect to the asymptotic, long-time, stationary value.}
$|\psi(x,t)|^2$  exhibits diffraction in
time, 
namely, a characteristic oscillation in time, and also in the 
space domain,  
see Fig. \ref{prt}, in contrast with the simple step function 
for a spatially homogeneous ensemble of classical particles
with momentum $p$ released at time $t=0$.    
%
%
%
%
%
%
Thanks to the relation between the $w$-function and Fresnel integrals
\cite{Abram},
\beqa
C(\theta)+iS(\theta)&=&\frac{1+i}{2}
\left[1-e^{i\pi\theta^2/2}w\left(\frac{1+i}{2}\pi^{1/2}\theta\right)\right],
\nonumber
\\
\theta&=&\left(\frac{t}{m\hbar\pi}\right)^{\!\!\!1/2}\!\!\!(p-mx/t)
=\frac{1-i}{\pi^{1/2}}u(p,t),
\nonumber
\eeqa
the solution may also be rewritten as
\begin{equation}\label{eq:Cor}
\vert\psi(x,t)\vert^2=P(\theta)=
\frac{1}{2}\left\{\left[S(\theta)+\frac{1}{2}\right]^2
+\left[C(\theta)+\frac{1}{2}\right]^2\right\},
\end{equation}
and therefore be mapped onto the Cornu espiral,
which  is plotted in Fig. \ref{Cornu}.
%
%
The ``density'' can then be read as half the distance from the point
$(-1/2,-1/2)$ to any other point of the spiral; and the origin corresponds 
to the classical 
particle with momentum $p$ released at time $t=0$ from the shutter position. 

Suppose now that the shutter is closed suddenly at 
time $\tau$. The integrated ``density'' 
$N_+(\tau):=\int_0^\infty dx\,|\psi(x,\tau)|^2$ grows 
continuously with $\tau$ and may reach arbitrarily large values, obviously 
greater than one. Moreover, 
$|\psi(x,t)|^2$ is dimensionless, so $N_+$ cannot be
taken  
as a probability in any case.    
This poses the question of 
physically interpreting the mathematical results. Clearly we face analogous  
problems to interpret a plane wave or stationary scattering states,
which are not in Hilbert space. 
The solution is then found similarly, 
by assuming that $\psi(x,t)$, except for a normalization factor,  
is just a component 
of a normalizable state in Hilbert space, to be determined by the 
experimental preparation setup.  
A 
consistent 
interpretation for several -noninteracting- particles requires the appropriate 
quantum symmetrization \cite{Stevens84},
which will be treated elsewehere. In the present 
paper we 
shall limit ourselves to the simplest case an consider only one-particle 
states.

For a shutter with a non-zero reflectivity, the cut-off plane wave of 
opposite momentum has to be taken into account, so the initial 
state takes the form
\beq\label{cutplanew}
\psi^{(\mathcal{R})}_{p}(x,t=0):=
(e^{i p/\hbar x}+\mathcal{R}e^{-i p/\hbar x})\Theta(-x).
\eeq
The cases $\mathcal{R}=\{1,0,-1\}$ turn out to be the 
most relevant.
We will refer to $\mathcal{R}=-1$ as $\emph{sine}$ initial conditions, 
which arise 
when the shutter is totally reflective; and to $\mathcal{R}=1$ 
as $\emph{cosine}$ 
initial conditions. This last case may seem to be hard to implement but, 
surprisingly, it is related to the usual \emph{source} boundary condition,
\beq
\label{sour}
\psi_p^{s}(x=0,t)=e^{-i\om t}\Theta(t),\;\; \psi_p^{s}(x>0,t\le 0)=0,
\eeq
where $\omega=p^2/(2m\hbar)$,
and is more tractable mathematically than the $\emph{sine}$ case. 
As we shall see, $\psi^s_p(x,t)=\psi_p^{(1)}(x,t)$.  
%
%
%
%
%
%
%
%
%
\section{Visibility of main diffraction fringe\label{vis}}
%
%
%
%
%
%
%
%

Let us examine first the time evolution from a cut-off plane wave
initial condition, i.e., $\mathcal{R}=0$, as in Eqs. (\ref{mos}) and 
(\ref{eq:Cor}).
The motion of points of constant ``density'' is obtained by imposing 
a constant value of 
$\theta$, according to Eq. (\ref{eq:Cor}). 
In particular, for the maximum and nearby minimum, see Fig. \ref{Cornu}, 
\beqa
x_{max}(t)&=&p t/m-\sqrt{\frac{\pi\hbar t}{m}}\theta_{max},
\nonumber\\
x_{min}(t)&=&p t/m-\sqrt{\frac{\pi\hbar t}{m}}\theta_{min},
\nonumber
\eeqa
where $\theta_{max}=1.217$ and $\theta_{min}=1.872$ are
universal for $\mathcal{R}=0$. 
Since the probability is exclusively $\theta$-dependent, 
see Eq. (\ref{eq:Cor}),
$P_{max}=1.370$, and $P_{min}=0.778$, independently of time, mass 
or momentum,  and therefore the fringe
visibility, defined as
$$
\mathcal{V}(T)=
\frac{P_{1^{st} max}-P_{1^{st} min}}{P_{1^{st} max}+P_{1^{st} min}},
$$
is concluded to be universal 
(equal to 0.276 for $\mathcal{R}=0$), and of course time-independent.

Other important fringe feature is the width of the main peak. It can be computed
from the intersection with the classical probability density 
(for a cut-off beam of particles 
released at time $0$ with momentum $p$) 
\cite{Moshinsky52,Moshinsky76},
$$
\Delta x\simeq 0.85 \left(\frac{\pi\hbar t}{m}\right)^{1/2}.
$$
Next, we consider an initial state given by a statistical mixture,
\beq
\label{rho}
\rho(t=0):=\int dp f(p) |\psi_p^{(\mathcal{R})}(t=0)\ra\la
\psi_p^{(\mathcal{R})}(t=0)|,
\eeq
and focus our attention on the effect
of temperature and evolution time on the diffraction pattern.  
Let us assume, by   
integrating out the y,z-degrees of freedom of a Maxwell-Boltzmann 
momentum distribution \cite{MS99}, that the Gaussian 
momentum probability density
\beq
f_{MB}(p)=\frac{1}{\sqrt{2\pi m k_{B}T}}\;e^{-\frac{(p-p_{c})^2}{2 m k_{B}T}}, 
\label{fpp}
\eeq
holds for the momentum in $x$ direction. 
The introduction of a non-zero mean average momentum $p_c$ implies that
the atomic ensemble, 
say in a magneto-optical trap, is launched along the $x$-axis with a 
velocity $p_c/m$.
This is realized, for example, in the ``moving molasses technique'' \cite{Salomon} 
within an atomic fountain clock. We shall generally assume 
that the contribution of negative $p$ in Eq. (\ref{fpp}) is negligible. 

\begin{figure}
\includegraphics[height=6cm,angle=-90]{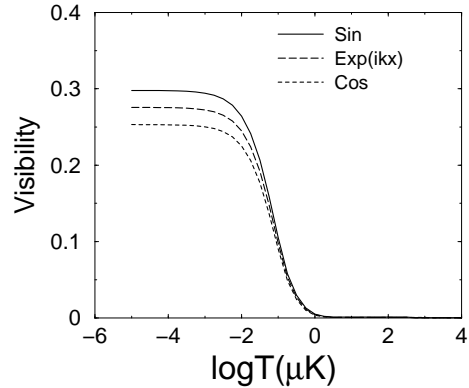}
\vspace*{.2cm}\\
\caption{\label{chvis} 
Visibility as a function of temperature and the
reflectivity of the shutter for a beam of Argon
with the distribution in Eq. (\ref{fpp}).  $p_{c}/m=10$ cm/s, $t=20\,\mu$s.
}
\end{figure}

Figure \ref{chvis} shows the effect of the temperature 
on the visibility of 
the main diffraction 
fringe for statistical mixtures weighted by the distribution of Eq. (\ref{fpp}), 
with 
$\mathcal{R}=\{0,\pm1\}$. 
%
%
%
%
It is clear that the diffraction peaks are more visible for  
$\emph{sine}$ conditions than for the isolated cut-off plane wave
and $\emph{cosine}$ conditions, 
the difference being more noticeable at small temperatures.  
As expected, there is a suppression of the 
diffraction pattern when the temperature increases. 
A possible criterion for this suppression is a
small value of the ratio, $\alpha$,  
between the width of the fringe $\Delta x$ and the separation
$\Delta x_{cl}$ of the classical arrival 
points corresponding to fast
and slow components of the statistical mixture,  
namely, 
\begin{displaymath}
\alpha:=\frac{\Delta x}{\Delta x_{cl}}=\frac{0.85}{\Delta p}
\sqrt{\frac{m\pi\hbar}{t}}.
\end{displaymath}
For the Gaussian distribution, Eq. (\ref{fpp}),  
one can choose $\Delta p$ 
as the full width at half maximum (FWHM), that is, 
$2\sqrt{2ln2 m k_{B}T}$, so that   
\begin{displaymath}
\alpha=0.36\sqrt{\frac{\pi\hbar}{K_{B}T t}}.
\end{displaymath}
%
\begin{figure}
\includegraphics[height=6cm,angle=-90]{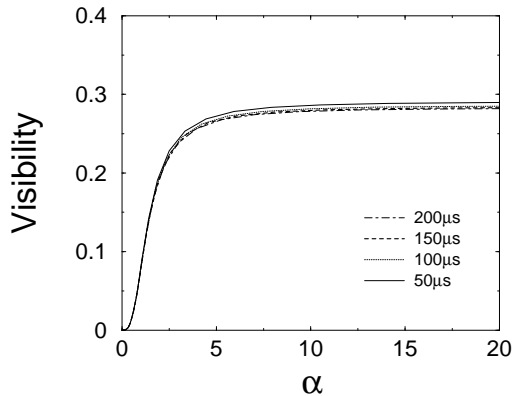}
\vspace*{.2cm}\\
\caption{\label{visa} 
Visibility as a function of the $\alpha$-parameter
for an Argon atom with a velocity $p_c/m=10$ cm/s.
Each curve is computed by varying
the temperature for a given opening time. $\mathcal{R}=-1$.}
\end{figure}
%
According to Fig. \ref{visa}, the visibility 
grows with $\alpha$ for small values
and saturates for large $\alpha$, i.e. for small times and/or temperatures. 

Moreover, if instead of Eq. (\ref{fpp}), the momentum distribution of an effusive 
atomic beam is used \cite{Ramsey,MS99}, 
\beq
f_{beam}(p)=\frac{p^{3}}{2(m k_{B}T)^{2}}\,e^{-\frac{p^{2}}{2 m k_{B}T}}\Theta(p),
\label{beam}
\eeq
then $\la p\ra=\sqrt{9\pi m k_{B}T/8}$, and the upshot is a complete 
supression of the diffraction in time phenomena for all times and temperatures. 
This fact points out the relevant role played
by the momentum distribution, that is, by the experimental
preparation setup being considered.

Summing it up, the fringe visibility, at variance with the pure state 
case, becomes time-dependent for mixed states. The 
characteristic oscillations of diffraction in time tend to be washed out with the
observation time after opening the shutter, and also by increasing the
temperature, 
or for the momentum distribution of an effusive beam.

%
%
%
%
%
%
%
%
\section{Evolution of a finite pulse\label{evol}} 
In this section we focus on the formation and evolution 
of a matter-wave, single-particle pulse.
Suppose that the shutter is opened at the instant $t=0$ and closed at 
the instant $\tau$. 
We will refer to the duration $\tau$ as 
the ``opening time''. It is, in other words, 
the width of the the time slit. 
The notation $\psi^{(\mathcal{R})}_{p_0,\tau}(t)$ will denote 
the state formed in that manner, at time $t\ge \tau$, from 
the initial state of Eq. (\ref{cutplanew}) with main momentum
$p_0$.\footnote{Another way to form a pulse 
is to remove at time $t=0$ the potential walls that confine 
a state which is initially stationary. For an infinite-wall 
square box the result is the substraction of two semi-infinite 
${\mathcal{R}}=-1$ pulses, i.e., at least for free motion, a simple
combination of $w$-functions. For applications of this idea in 
free-motion cases or scattering 
configurations see \cite{GK76,TKF87,Stevens84,BEM01,Godoy02}.}

For an initially pure state the time evolution of the pulse 
can be written in terms of 
the inner products with the eigenstates of the free hamiltonian which vanish
at the origin,  
$$
\la x\vert\phi_{E}\ra
=\sqrt{\frac{2 m \hbar}{\pi p}}\sin{(p x/\hbar)},\;\;\;
E=p^2/(2m).   
$$
They obey orthonormality and completeness relations,
\begin{displaymath}
\la\phi_{E'}\vert\phi_{E}\ra=\delta(E-E'),\;\;
\\
\int_{0}^{\infty} dE\,\vert\phi_{E}\ra\la\phi_{E}\vert=\openone.
\end{displaymath}
As shown by Moshinsky 
\cite{Moshinsky76}, the overlap between these eigenstates and the 
state that results from a state formed after an opening time $\tau$ 
with $\mathcal{R}=0$ is  
\beqa
\label{over}
&&\la\phi_{E}\vert\psi_{p_{0},\tau}^{(0)}(t=\tau)\ra
=\sqrt{\frac{m p\hbar}{2\pi}}
\nonumber
\\
&\times&\Bigg\{\!\frac{w[-u_{0}(p_{0},\tau)]}{p^{2}-p_{0}^{2}}
\!-\!\frac{w[-u_{0}(p,\tau)]}{2p(p-p_{0})}\!-\!
\frac{w[-u_{0}(-p,\tau)]}{2p(p+p_{0})} 
\Bigg\},
\nonumber
\\
\eeqa
where
$$
u_{0}(p,\tau):=\frac{1+i}{2}\sqrt{\frac{\tau}{m \hbar}}\,p,
$$
to be compared with Eq. (\ref{upt}). ($u_0(p,\tau)$ is a particular  
case of $u(p,\tau)$ in which $x=0$.)
The overlaps for the cases $\mathcal{R}=\pm 1$ are obtained 
easily from Eq. (\ref{over}). 

Writing the final time of observation as
$t=\tau+\Delta t$, the evolved pulse can
then be calculated as
\beq
\psi^{(\mathcal{R})}_{p_{0},\tau}(x,t)=
\int_{0}^{\infty}
dE\, e^{-iE\Delta t/\hbar}\la x\vert\phi_{E}\ra\la\phi_{E}\vert
\psi^{(\mathcal{R})}_{p_{0},\tau}(t=\tau)\ra.
\label{eqevol}
\eeq
This expression is not well-behaved numerically,
but such a problem can be swiftly solved  by modifying the contour of
integration in the complex $p$-plane, since several terms can be written 
as $w$-functions (see the Appendices A and B for details), 
and for the rest we gain an integrand with Gaussian decay. Explicitly,
for $\mathcal{R}=-1$, 
{\setlength\arraycolsep{2pt}
\begin{eqnarray}\label{sinecase}
&&\psi^{(-1)}_{p_{0},\tau}(x,t)  =  
\frac{e^{\frac{i m x^{2}}{2 t \hbar}}}{2}\,
\left\{[w[-u(p_{0},t)]+w[-u(-p_{0},t)]\right\} 
\nonumber
\\
& - & \frac{e^{-i \frac{p_{0}^{2}\tau}{2 m\hbar}}-w[u_{0}(p_{0},\tau)]}{2}
e^{\frac{i m x^{2}}{2 \Delta t \hbar}}
\nonumber\\
&\times&\left\{w[-u(p_{0},\Delta t)]+w[-u(-p_{0},\Delta t)]\right\}
\nonumber
\\
\! &\! -\! &\!\frac{i}{2\pi }\int_{\Gamma_{+}}\!\!\!dp\,
e^{-i \frac{p^{2}\tau}{2 m\hbar}+i px/\hbar}
\nonumber
\\
&\times&
\left\{\frac{w[u_{0}(-p,\tau)]}{p+p_{0}}+
\frac{w[u_{0}(p,\tau)]}{p-p_{0}}\right\},
\end{eqnarray}
}
where the contour passes above the poles. 
This result is so far exact.  
The integral term, $\mathcal{Y}$, can be rewritten by
integrating in the $u$ variable,
\beqa
\mathcal{Y}&=&-\frac{i  e^{\frac{i m x^{2}}{2 \Delta t \hbar}}}{2\pi}
\int_{\Gamma_{u}}du\,{e^{-u^{2}}}
\nonumber\\
&\times&
\left[\frac{w(\sqrt{\frac{\tau}{\Delta t}}u+v)}{u-u(p_{0},t)}+
\frac{w(-\sqrt{\frac{\tau}{\Delta t}}u-v)}{u-u(-p_{0},t)}\right],
\eeqa
see the Appendix A, 
with 
$$
v=\frac{1+i}{2}\sqrt{\frac{\tau}{m\hbar}}\frac{m x}{\Delta t}.
$$
A way of dealing with this kind of integral to excellent accuracy 
was discussed by Brouard and one the authors in \cite{BM96}.
It  consists in extracting the singularities of the
integrand, which we shall denote as $g(u)$, in the following way 
$$
g(u)=\sum_{i}\frac{A_{i}}{u-u_{i}}+h(u).
$$
$A_{i}$ is the residue of $g(u)$ at the pole $u_{i}$  
and $h(u)$ is an entire function which admits a power series expansion 
convergent in the whole $u$-plane,
\beqa
&&\int_{-\infty}^{\infty}h(u)e^{-u^{2}}du
\nonumber
\\
&=&
\sqrt{\pi}\left[h(u=0)+\sum_{n=1}^{\infty}\frac{(2n-1)!!}{2^{n}(2n)!}
\, h^{(2n)}(u=0)\right].
\nonumber
\eeqa
Retaining just the first term of the previous expansion 
leads to
$$
\int_{\Gamma_{u}}du\, g(u)\simeq\sum_{i} A_{i}
\left[-i\pi w(-u_{i})+\frac{\sqrt{\pi}}{u_{i}}\right]
+\sqrt{\pi}g(u=0).
$$
In particular,
\beqa
\label{in}
\mathcal{Y}&\simeq&\!-\frac{e^{\frac{i m x^{2}}{2 \Delta t \hbar}}}{2}
\left\{w[u_{0}(p_{0},\tau)]w[-u(p_{0},\Delta t)]\right.
\nonumber\\
&+&\!
\left.w[-u_{0}(p_{0},\tau)]w[-u(-p_{0},\Delta t)]\right\}
\nonumber\\
&\!+\!&\!\!\frac{ie^{\frac{i m x^{2}}{2 \Delta t \hbar}}}{2\sqrt{\pi}}
\!\left\{\!\frac{w(v)\!-\!w[u_{0}(p_0,\tau)]}{u(p_{0},\Delta t)}\right.
\nonumber
\\
&+&\left.\frac{w(-v)\!-\!w[-u_{0}(p_0,\tau)]}{u(-p_{0},\Delta t)}\!\right\}\!,
\eeqa
which we shall use later on in Sec. \ref{compa} to compare the initial 
value problem and source approaches. 
\subsection{Purity of the chopped state: coherence creation 
from incoherent mixtures by using small opening times}
\begin{figure} 
\includegraphics[height=6cm,angle=-90]{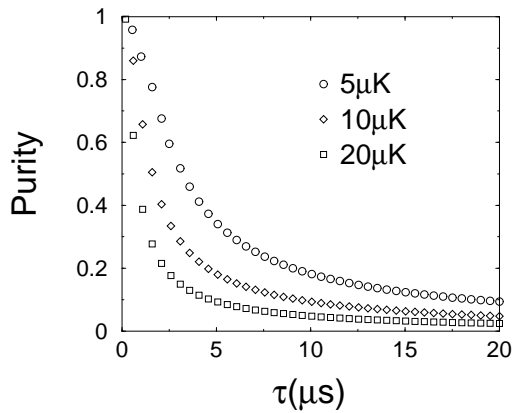}
\vspace*{.2cm}\\
\caption{\label{purity}
Variation of purity versus the opening time for different temperatures.
An effusive beam of Argon atoms is considered, see Eq. (\ref{beam}).
$\mathcal{R}=0$.}
\end{figure}

It is possible to ``create coherence'' from statistical mixtures, 
simply by means of small opening times. This was seen 
experimentally with cold atoms by Dalibard and coworkers 
\cite{SGAD96}. 
To quantify this effect we use 
the purity, defined by  
\begin{displaymath}
\mathcal{P}_{t}:=\tr\rho_{N,\tau}^{2}(t), 
\end{displaymath}
where $\rho_{N,\tau}(t)$ is the normalized density operator at time $t$
for an opening time $\tau$.   
Our starting point is the unnormalized state 
\beq\label{rhott}
\rho_{\tau}(t):=\int dp f(p)\vert 
\psi^{(0)}_{p,\tau}(t)\ra\la\psi^{(0)}_{p,\tau}(t)\vert.
\eeq
%
Thanks to the invariance 
under cyclic permutation of the trace and unitarity of the time evolution,
the purity is invariant for $t\ge\tau$, so it can be calculated 
in particular at $t=\tau$,  
%
%
%
$$
\mathcal{P}_\tau=\frac{1}{N^{2}}\int\!\!\int  
dp dp'f(p)f(p')\vert\la\psi^{(0)}_{p,\tau}\vert\psi^{(0)}_{p',\tau}\ra\vert^{2},
$$
where $N$ is given by
$$
N={\tr}\rho_\tau=
\int dp f(p)\la\psi^{(0)}_{p,\tau}\vert\psi^{(0)}_{p,\tau}\ra.
$$
Considering the case of an effusive atomic beam, $f(p)=f_{beam}(p)$, Eq. 
(\ref{beam}), Fig. \ref{purity} shows 
that the purity ${\mathcal{P}}_t$
decreases as a  function of the opening time and temperature.
For short opening times a
highly coherent state is created in spite of the 
mixed preparation state. 
%
%
%
%
%
This result, in the limit $\tau\to 0$, may be understood as follows: 
As long as a (pure state) wavefunction 
vanishes in $x>0$ at time $t=0$, the wavefunction
for $t>t'>0$ can be written as \cite{BEM01}
$$
\psi(x,t)=\int_{0}^{t}dt' K_{+}(t,x;t',x'=0)\psi(x'=0,t'),
$$
where
\begin{eqnarray}
K_{+}(t,x;t',x'=0)  &=&  
\frac{\hbar}{m i}\frac{d}{d x}{\mathcal{G}}_0(x,t;0,t')
\nonumber
\\
& = & \left[\frac{m}{i h (t-t')^{3}}\right]^{1/2} x e^{imx^{2}/2\hbar(t-t')}.
\nonumber
\end{eqnarray}
%
%
%
By performing a Taylor
series expansion in $t'$ around $t'=0$, one finds that
$$
\psi(x,t)=t K_{+}(t,x;0,0)+
\mathcal{O}(t^{2})
$$
for $\psi(x=0,t'=0)=1$. Since $\psi^{(0)}_{p,\tau}(x=0,t'=0)=1$ 
for all $p$, then  
to first order all the momenta that form the preparation mixture 
give the same contribution, in other words, 
$|\psi_{p,\tau}(t)\ra$ in Eq. (\ref{rhott}) becomes independent of $p$, 
and thus the resulting state 
tends to be independent of the initial distribution 
$f(p)$, and totally coherent (pure) in the limit of small opening times.  
For the case of the Gaussian distribution in Eq. (\ref{fpp}),
the same 
qualitative behaviour (monotonic decay from $1$ in the same time scale)
is observed.  
\section{Comparison between ``source'' and
``initial value 
problem'' approaches\label{compa}}
Golub and G\"ahler \cite{GG84} tackled the diffraction
in time problem in two dimensions using the
Green's function formalism, working within the Fraunhoffer limit, and 
imposed source 
boundary conditions. This approach was later thoroughly implemented for different
combinations of time and space rectangular slits by 
Brukner and Zeilinger 
\cite{BZ97}. Other works  have used the simple source boundary conditions
of Eq. (\ref{sour})   
in one dimension, see \cite{MB00}, and references therein. 
In \cite{BEM01} the relations between source boundary conditions 
and the initial value problem (IVP) were examined and equivalence conditions 
established. 

\begin{figure} 
\includegraphics[height=6cm,angle=-90]{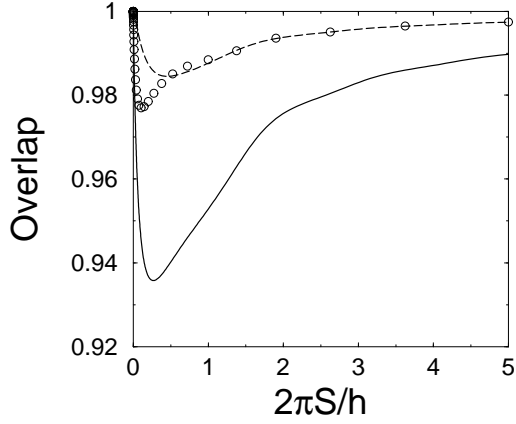}
\vspace*{.2cm}\\
\caption{\label{chexvssbc2}
Overlap probability between the wavefunction of one Argon atom associated with 
$\emph{sine}$ and $\emph{cosine}$ initial conditions (continuous line),
cut-off plane wave and sine conditions (circles),
and cut-off plane wave and $\emph{cosine}$ conditions (dashed line),  
as a function of the phase $S/\hbar:=\tau p_0^2/(2 m\hbar)$,
see Eq. (\ref{overlap}).
}
\end{figure}

In order to compare pulses formed by  source or IVP approaches 
let us 
introduce the window function
$\chi_{[0,\tau]}(t)=\Theta(t)\Theta(\tau-t)$ and impose the source boundary 
condition
\beq \label{eq:sbc}
\psi^{\,s}_{p_{0},\tau}(x=0,t)= e^{-i \omega_{0} t}\chi_{[0,\tau]}(t),
\eeq
(the wave function is also assumed to vanish at $x>0$ for all $t\le 0$) 
with  Fourier transform
\begin{displaymath}
\widehat{\psi}^{\,s}_{p_{0},\tau}(x=0,\omega)=
\frac{i}{\sqrt{2\pi }}\, 
\frac{1}{\omega-\omega_{0}}\big[1-e^{i(\omega-\omega_{0})\tau}\big].
\end{displaymath}
At a later time one finds, for $x>0$, 
\begin{displaymath}
\psi^{\,s}_{p_{0},\tau}(x,t)=\frac{i}{2\pi }\int_{-\infty}^{\infty}
d\omega\frac{e^{i p x/\hbar-i\omega t}}{\omega-\omega_{0}}
\big[1-e^{i(\omega-\omega_{0})\tau}\big],
\end{displaymath}
where ${\rm{Im}} p\ge 0$.   
This can be written in the complex $p$-plane by introducing the contour of
integration $\Gamma_{+}$ which goes from 
$-\infty$ to $\infty$ passing above the poles, 
{\setlength\arraycolsep{2pt}
\begin{eqnarray}
&&\psi^{\,s}_{p_{0},\tau}(x,t)  =  \frac{i}{2\pi }
\int_{\Gamma_{+}}dp\,2 p\,\frac{e^{i p x/\hbar-i
\frac{p^{2}t}{2 m \hbar}}}{p^{2}-p_{0}^{2}}
\big[1-e^{i(\omega-\omega_{0})\tau}\big]
\nonumber\\
& = &\frac{i}{2\pi }\int_{\Gamma_{+}}\!\!\!dp\, 
\left(\frac{1}{p+p_{0}}\!+\!\frac{1}{p-p_{0}}\right)
e^{i p x/\hbar-i \frac{p^{2}t}{2 m \hbar}}
\big[1\!-\!e^{i(\omega-\omega_{0})\tau}\big].
\nonumber
\end{eqnarray}
}
The integral can be carried out exactly using Appendix A, 
\begin{eqnarray}
\label{eq:cos}
&&\psi^{\,s}_{p_{0},\tau}(x,t) 
= \frac{e^{\frac{i m x^{2}}{2 t \hbar}}}{2}\, 
\{w[-u(p_{0},t)]+w[-u(-p_{0},t)]\}
\nonumber\\
\!\!&\!\!&\!\! -\frac{e^{-i \frac{p_0^{2}\tau}{2 m\hbar}
+\frac{i m x^{2}}{2 \Delta t \hbar}}}{2}\!
\{w[-u(p_{0},\Delta t)]\!+\!w[-u(-p_{0},\Delta t)]\},
\nonumber
\\
\label{sou}
\end{eqnarray}
so the time evolution is straightforward and given by a difference of
$w$-functions
(in fact, expressing the window function as a difference of step 
functions $\chi_{[0,\tau]}=\Theta(t)-\Theta(t-\tau)$,
the solution can be foreseen to be given by the 
difference of one $\emph{source}$ open at $t=0$ and a second one 
at a later time $t=\tau$, with the associated phase taken into account).
The nice thing about this approach is that it allows a clear and intuitive
interpretation of the different contributions to the total wavefunction,
and it is easy to implement complicated combinations of time and space
windows. The flip side is that deviations from the exact result 
of the initial value problem arise,
specially  for the back part of the wavefunction cut by the chopper, 
except in the case $\mathcal{R}=1$. 
Although the calculation is somewhat tedious -see Appendix C-,
it can be proved that for $\emph{cosine}$ initial conditions,
%
%
the wave function that results from a shutter opened between $0$ and $\tau$, 
$\psi_{p_0,\tau}^{(1)}(x,t)$,  
is $\emph{exactly}$ equivalent to using source boundary conditions 
in that interval, i.e., to  
Eq. (\ref{eq:cos}).
The difference with the solution for the reflecting shutter ($\mathcal{R}=-1$) 
can be written as
\begin{eqnarray}
&&\psi^{(-1)}_{p_{0},\tau}(x,t)-\psi^{(1)}_{p_{0},\tau}(x,t)
\simeq\frac{e^{\frac{i m x^{2}}{2 \Delta t \hbar}}}{2}
\nonumber\\
&\times&\left\{w[u_{0}(p_0,\tau)]-w[-u_{0}(p_0,\tau)]\right\}
w[-u(-p_0,\Delta t)]
\nonumber\\
& + &\!\frac{ie^{\frac{i m x^{2}}{2 \Delta t \hbar}}}{2\sqrt{\pi}}
\left(\frac{w(v)-w[u_{0}(p_0,\tau)]}{u(p_{0},\Delta t)}\right.
\nonumber\\
&+&\left.\frac{w(-v)\!-\!w[-u_{0}(p_0,\tau)]}{u(-p_{0},\Delta t)}\right),
\end{eqnarray}
where we have used Eqs. (\ref{sinecase}), (\ref{in}) and (\ref{eq:cos}). 
The overlap probability between pulses generated  
for different values of $\mathcal{R}$ and fixed $\tau$ and $p_0$,
\begin{equation}
{\mathcal{O}}^{(\mathcal{R,R'})}:=
\frac{\vert\la\psi^{(\mathcal{R})}\vert\psi^{(\mathcal{R}')}
\ra\vert^{2}}{\la\psi^{(\mathcal{R}')}
\vert\psi^{(\mathcal{R}')}\ra\la\psi^{(\mathcal{R})}\vert\psi^{(\mathcal{R})}\ra},
\label{overlap}
\end{equation}
is independent of the evolution time 
$t$, and can be calculated exactly 
at $t=\tau$. It turns out to be a 
function of the adimensional phase
$ S/\hbar:=\tau p_{0}^{2}/(2 m \hbar)$, see Fig. \ref{chexvssbc2}.   
%
%
The overlaps tend to one
as $\tau\to 0,\infty$.  
Remarkably, there is a lower bound which is 93.56\% between $\emph{sine}$ 
and ${\emph{cosine}}-$source conditions. 
We may conclude that in general the result of using 
source boundary conditions is not at all 
a bad approximation to the solution of the initial value problem with 
reflecting or absorbing conditions, except when very 
accurate results are required.

%
%
%
%
%
%
%
\section{Apodization of matter-wave pulses\label{apo}}
Apodization is a technique long used in light optics to avoid
the diffraction effect in several devices. Simply put,
it consists in suppressing high frequency components by smoothing the
aperture function (or ``window'' for short) of the pulse \cite{Fowles68}.
The price to pay is a slight broadening of the energy distribution.
In this section such a technique is
applied to matter wave pulses within the source approach. 
In particular, the case of a sine type aperture function can be analytically
solved in terms of $w$-functions.
Let us consider in Eq. (\ref{eq:sbc}), instead of the
sudden, rectangular shape, a sine aperture function,
\beq
\chi_{[0,\tau]}(t)=\sin{(\Omega t)}\,\Theta(t)\,\Theta(\tau-t)
\label{sin}
\eeq
with $\tau=\pi\Omega^{-1}$. (Note that the ``sine'' aperture function
has nothing to do with $\emph{sine}$ initial conditions.) 
Then, defining $\om_{\pm}=\om_{0}\pm\Om$ and the corresponding momenta 
$p_{\pm}=\sqrt{2m\hbar\om_{\pm}}$, with the branch cut taken along the negative
imaginary axis of $\om_{\pm}$, it is found that
\begin{displaymath}
\widehat{\psi}^{\,s}_{p_0,\tau}(x=0,\omega)
=\frac{1}{\sqrt{8\pi}}\, \left[\frac{1-e^{i(\om-\om_{-})\tau}}{\om-\om_{-}}
-\frac{1-e^{i(\om-\om_{+})\tau}}{\om-\om_{+}}\right], 
\end{displaymath}
and the wavefunction is
(we shall distinguish apodized pulses from the rectangular 
aperture pulses used so far by means of a tilde) 
\beqa
\!&\!&\!\widetilde{\psi}^{\,s}_{p_0,\tau}(x,t)
=\frac{-1}{2 \pi }\!\sum_{\alpha=\pm}
\alpha
\nonumber
\\
&\times&\int_{-\infty}^{\infty}
\!\!d\om\, e^{-i\om t+i p x/\hbar}
\left[\frac{1-e^{i(\om-\om_{\alpha})\tau}}{\om-\om_{\alpha}}\right]
\nonumber
\\
&=&\frac{-1}{2 \pi }\!\sum_{\alpha=\pm}\!\alpha\!
\int_{\Gamma_{+}}\!\!\! d p\, e^{i p x/\hbar} \Big(\!e^{-i\om t}
-e^{-i\om\Delta t-i\om_{\alpha}\tau}\!\Big)
\nonumber
\\
&\times&\Bigg(\frac{1}{p\!+\!p_{\alpha}}\!+\!\frac{1}{p\!-\!p_{\alpha}}\!\Bigg)
\nonumber
\\
& = &\!\frac{i}{2}\!\sum_{\alpha=\pm}\!
\alpha\!\Bigg(\!e^{\frac{i m x^{2}}{2 t \hbar}}\left\{w[-u(p_\alpha,t)]
\!+\!w[-u(-p_{\alpha},t)]\right\}
\nonumber
\\
&-&\!
e^{-i \frac{p_{\alpha}^{2}\tau}{2 m\hbar}\!+\!
\frac{i m x^{2}}{2 \Delta t \hbar}}\!\left\{w[-u(p_{\alpha},\Delta t)]\!+\!
w[-u(-p_{\alpha},\Delta t)]\right\}\Bigg)
\nonumber
\\
& = & i\sum_{\alpha=\pm}\alpha\,\,\psi_{p_{\alpha },\tau}^{\,s}(x,t).
\nonumber
\eeqa
As it is clear from the last line, the apodization with
the sine aperture function is equivalent to the introduction 
of two different momenta components whose values are related
to the length of the pulse. Pictorically, a sine window pulse is equivalent 
to the coherent difference 
of two simultaneous pulses produced with sudden rectangular windows 
from speeded up
and decelerated sources. 
%
%
%

\begin{figure} 
\includegraphics[height=6cm,angle=-90]{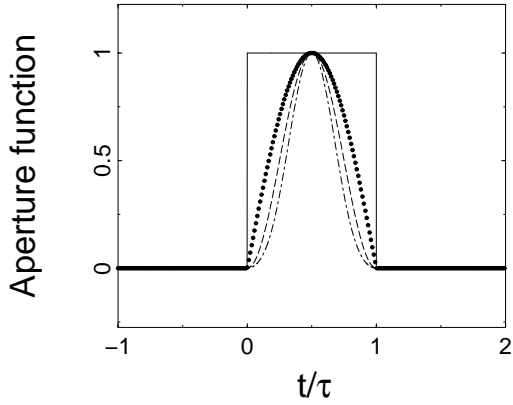}
\vspace*{.2cm}\\
\caption{\label{apofunc}
Different aperture function studied. The smoothness at the edges 
increases in this order: rectangular (continuous line), 
sine (dotted line), Hanning (dashed) and Blackman (long dashed line). 
}
\end{figure}

Similarly, other window functions,
as the ones shown in Fig.
\ref{apofunc} can be worked out.
For instance, the Hanning aperture function, 
%
\beqa\label{Hanning}
\chi_{[0,\tau]}(t)&=&\sin^{2}({\Om t})\,\Theta(t)\,\Theta(\tau-t)
\nonumber\\
&=&\frac{1}{2}(1-\cos{2\Omega t})\,\Theta(t)\,\Theta(\tau-t),
\eeqa
after introducing $p_{\beta}=\sqrt{2m\hbar(\om_{0}\pm2\Om)}$ 
with $\beta=\pm$, leads to 
$$
\widetilde{\psi}^{\,s}_{p_0,\tau}(x,t)=
\frac{1}{2}\left[\psi_{p_{0},\tau}^{\,s}(x,t)
-\frac{1}{2}\sum_{\beta=\pm}\psi_{p_{\beta},\tau}^{\,s}(x,t)\right],
$$
where the effect of the apodization is to subtract to the pulse
with the source momentum
two other matter-wave trains associated with $p_{\beta}$, all of 
them formed with rectangular aperture functions. 

\begin{figure}
\includegraphics[height=7cm,angle=-90]{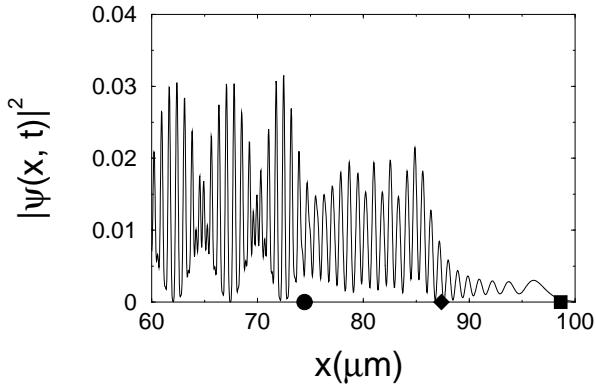}
\vspace*{.2cm}\\
\caption{\label{phan} 
Front part of the probability density for the periodic Hanning function
at $t=1.2$ ms, 
with an aperture time $\tau=10 \mu$s.
The circle, diamond and square mark respectively 
the classical 
position of the slow $(p_-)$, original $(p_0)$ and accelerated $(p_+)$
components. $p_0/m=10$ cm/s.  
}
\end{figure}

A related configuration wich can be solved analytically in
terms of $w$-functions is the source apodized periodically to
create successive pulses. This can be done extending Eq. (\ref{Hanning}) 
to all positive times.
The result
is
\beqa
&&\widetilde{\psi}^{\,s}_{p_0,\tau}(x,t)=
\frac{e^{\frac{i m x^{2}}{2 t \hbar}}}{4}\left\{w[-u(p_{0},t)]+w[-u(-p_{0},t)]
\right\}
\nonumber
\\
&-&\frac{1}{8}\sum_{\beta=\pm} e^{\frac{i m x^{2}}{2 t \hbar}
}\left\{w[-u(p_{\beta}, t)]+
w[-u(-p_{\beta}, t)]\right\}.
\nonumber
\eeqa 
Note again the intervention of three different momenta. 
A consequence is that the $w$-function for the fastest component will 
eventually separate spatially as a smoother  -rather than pulsed- 
forerunner, see Fig. \ref{phan}. 
%

Also suitable for suppression 
of the sidelobes is 
the so-called 
Blackman aperture function,  
%
%
$$
\chi_{[0,\tau]}(t)=0.42-0.5\cos(2\Om t)+0.08\cos(4\Om t). 
$$
%
%
%
Using the $p_{\beta}$ defined above and the momenta
$$
p_{\gamma}=\sqrt{2m\hbar(\om_{0}\pm4\Om)},\;\,
$$
with $\beta,\gamma=\pm$, one can write the time evolved wavefuntion in terms 
of twenty $w$-functions,
\beqa
\widetilde{\psi}_{p_0,\tau}^{\,s}(x,t)&=&0.42 \psi^{\,s}_{{p_{0}},\tau}(x,t)
-0.25\sum_{\beta=\pm} \psi^{\,s}_{{p_{\beta}},\tau}(x,t)
\nonumber
\\
&+&0.04\sum_{\gamma=\pm} \psi^{\,s}_{{p_{\gamma}},\tau}(x,t),
\nonumber
\eeqa
so, once again, the smoothed shutter is related to a
combination of different pulses from rectangular time slits.

\begin{figure}
\includegraphics[height=6cm,angle=-90]{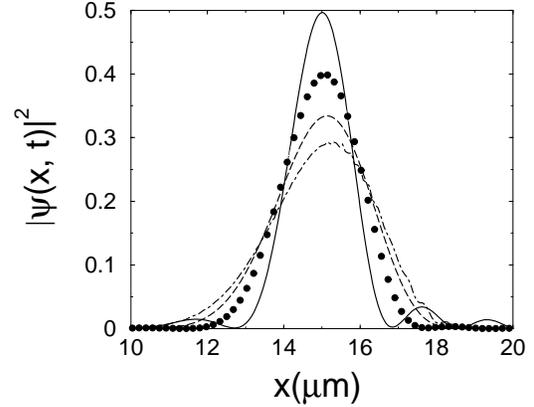}
\vspace*{.2cm}\\
\caption{\label{apowf} 
Probability density of a pulse of ultracold Argon atoms ($p_0/m=10$ cm/s), 
with $\tau=10\, \mu$s registered at $t=200\,\mu$s  for different window functions: 
rectangular (continuous line), sine (dotted line), 
Hanning (dashed line) and Blackman (dot-dashed line).
}
\end{figure}

\begin{figure}
\includegraphics[height=6cm,angle=-90]{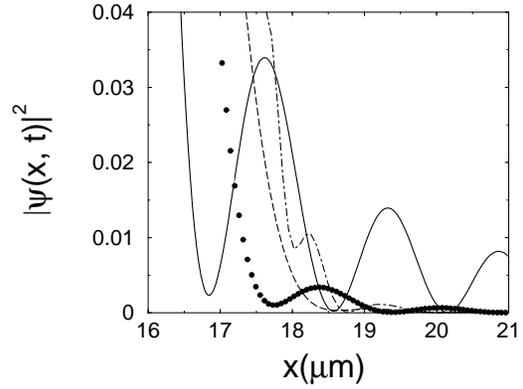}
\vspace*{.2cm}\\
\caption{\label{apowf2} 
Detail of the sidelobes of the previous figure. 
}
\end{figure}

The probability density for a single pulse and the details of the secondary 
diffraction peaks for four different aperture functions are 
illustated in Figs. \ref{apowf} and \ref{apowf2}, where all states are 
normalized to one. 
As expected, the smoother the time window function, the higher the 
suppression of the 
sidelobes.  
Note that the Blackman apodized pulse is slightly more
advanced that the other 
ones due to the faster components induced by this particular apodization. 
%
%
%
%
%
%
%
%
%
%
%
%
%
%
\subsection{Time-Energy uncertainty relation} 
A time energy uncertainty relation was discussed for the
shutter problem by Moshinsky,
who calculated the energy distribution of a particle after closing the shutter
\cite{Moshinsky76}.
Using $\emph{cosine}$ conditions with energy $E$ and taking the overlap of
the wavefunction with the free particle eigenstates which vanish at the origin, 
he calculated the energy distribution when the shutter was open a time $\tau$, 
$$
\wp(E,E',\tau)=\mathcal{N}\sqrt{E'}\,
\frac{\sin^{2}[(E-E')\tau/2\hbar]}{(E-E')^{2}},
$$
$\mathcal{N}$ being the normalization constant.
%
Then it was concluded that, for some measure of the energy width 
\cite{Moshinsky76} 
(see \cite{Busch02} for a comprehensive review of 
time-energy uncertainty relation), 
$$
\Delta E\,\,\tau\,\simeq\, \hbar.
$$
%
In fact, if one introduces the root of the 
variance $\sigma_{E}$ 
as the measure of energy spreading, $\Delta E=\sigma_{E}$,
such an uncertainty relation cannot be established since  
$\sigma_{E}$ diverges. This is in consonace with the 
non-existence of the average displacement, $\la x\ra$, 
for sharply cut waves \cite{MS05}.
One can calculate also the
energy distribution
for the case of a sine-aperture function and source boundary conditions,
Eq. (\ref{sin}), analytically. 
It turns out to be slightly broadened, but the variance is 
now well-defined as a consequence of the supression of high
frequency sidelobes.

\begin{figure}
\includegraphics[height=6cm,angle=-90]{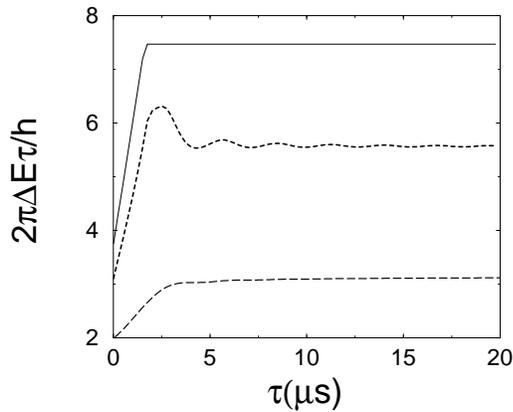}
\vspace*{.2cm}\\
\caption{\label{te} 
Uncertainty product, $\Delta E \tau$, for a pulse of ultracold 
Argon atoms ($p_0/m=10$ cm/s): sine aperture function and $\Delta E=FWHM$ 
(solid line); sine aperture function and $\Delta E=\sigma_{E}$
(dotted line); rectangular aperture function 
with $\Delta E=FWHM$ (dashed line). 
}
\end{figure}

Alternatively, as done in \cite{SGAD96}, the FWHM can be taken as
the measure of energy spread.
One observes that in both cases, sudden and smooth shutter,
both measures of the spread decrease 
with the opening time. 
The uncertainty product
is represented versus the opening time in 
Fig. \ref{te}. 
%
%
%
%
%
%
Note the two regimes, first a linear growth 
and then saturation, separated by the 
characteristic time required for a classical particle to travel 
a de Broglie wavelenght, 
$$
\top=\frac{h m}{p_{0}^{2}}.
$$
\section{Concluding remarks}
Starting with a brief review of the Moshinsky shutter problem, we have 
examined  several one-dimensional, matter-wave pulse characteristics
such as the visibility 
of the diffraction-in-time fringes 
versus evolution time and temperature, or the creation of coherence by small
aperture times. 
Moreover, a long standing missed comparison among several initial conditions
used in previous works, which differed on the 
reflected components in the preparation region, 
has been carried out. 
We have studied several apodizations of the aperture function,  
and obtained analytical expressions for the time evolution of the corresponding
pulses,
also in the periodic case.  
The effects of apodization to suppress secondary 
diffraction peaks and in the energy width have been discussed. 

The present work is the first step towards a more ambitious objective, namely, 
tayloring the pulses for particular needs, including the case in which 
interatomic interaction is important. Applications for laser atoms, 
in particular, will require in general to consider non-linear effects. 
Physically, optical shutters formed by effective potentials due to detuned lasers 
offer a 
unique oportunity, not available with mechanical shutters, 
to control the formation and subsequent behavior of the pulse.

\begin{acknowledgments}
We thank Gast\'on Garc\'\i a-Calder\'on for valuable comments. 
This work has been supported by Ministerio de Educaci\'on y Ciencia
(BFM2003-01003),
and 
UPV-EHU (00039.310-15968/2004).
AC acknowledges a fellowship from the Basque Government (BF104.479). 
\end{acknowledgments}


\begin{appendix}
\section{Integrals}

Consider the following integral along a contour which goes from $-\infty$ 
to $\infty$ passing 
above the pole at $p_{0}$,  
$$
\mathcal{I}=\int_{\Gamma_{+}} dp \frac{e^{-i a p^{2}+i b p}}{p-p_{0}},\;\;\;
a>0. 
$$
The saddle point is at $p_{s}=b/2a$.
By completing the square, one is lead to introduce the variable
$$
u=u(p)=\frac{1+i}{\sqrt{2}}\left(\sqrt{a}p-\frac{b}{2\sqrt{a}}\right). 
$$
Deforming the contour to go along the steepest descent path 
from the saddle, which in the $u$-plane
is at the origin,  
$$
\mathcal{I}=e^{i\frac{b^{2}}{4a}}\int_{\Gamma_{u}}du\frac{e^{-u^{2}}}{u-u(p_0)},
$$
If $\mathcal{C}_{0}$ denotes a counterclockwise circle around the pole $u(p_0)$,
then  $\Gamma_{u}=(-\infty,\infty)\cup\mathcal{C}_{0}$ if the pole has been crossed 
($Im(u)>0$) and the real line of $u$ otherwise.
We can inmediatly recognize this integral as a Faddeyeva 
function since the following  
equalities hold thanks to Cauchy's theorem, 
\beqa
&&\!\!\frac{1}{i\pi}\int_{\Gamma_{+}}du\frac{e^{-u^{2}}}{u-u(p_0)}
\nonumber\\
&=&\frac{1}{i\pi}\int_{\Gamma_{-}}du\frac{e^{-u^{2}}}{u-u(p_0)}-2e^{-[u(p_0)]^{2}}
\nonumber
\\
&=&w[u(p_0)]-2e^{-[u(p_0)]^{2}}=-w[-u(p_0)], 
\nonumber
\eeqa
where $\Gamma_+$ and $\Gamma_-$ go from $-\infty$ to $\infty$ passing 
above and below the pole, respectively. 
One concludes that
$$
\mathcal{I}=-i\pi e^{i\frac{b^{2}}{4a}}w[-u(p_0)].
$$
\section{
sine initial conditions}
After opening the shutter and closing it at time $\tau$,
the wavefunction for a reflecting shutter, ${\mathcal R}=-1$, 
is given by
\begin{displaymath}
\psi^{(-1)}_{p,\tau}(x,t)= 
\psi^{(0)}_{p,\tau}(x,t)-\psi^{(0)}_{-p,\tau}(x,t),
\end{displaymath}
which is a difference of $w$-functions.

Since, using Eqs. (\ref{over}) and (\ref{eqevol}), 
\beqa
&&\psi^{(0)}_{p,\tau}(x,t)
=\frac{1}{2\pi}\int_{-\infty}^{\infty}
dp' p'\,
e^{-i\frac{p'^{2}\Delta t}{2 m \hbar}}
\sin(p' x/\hbar) 
\nonumber\\
&\times&\!\!\Bigg\{\frac{w[-u_{0}(p,\tau)]}{p'^{2}-p^{2}}
-\frac{w[-u_{0}(p',\tau)]}{2p'(p'-p)}
-\frac{w[-u_{0}(-p',\tau)]}{2p'(p'+p)} \Bigg\}, 
\nonumber
\\
\label{psi0}
\eeqa
the wavefunction can be written as
\begin{eqnarray}
&&\psi^{(-1)}_{p,\tau}(x,t) 
 =  \frac{i}{8\pi }\int_{-\infty}^{\infty} dp' 
e^{-i\frac{p'^{2}\Delta t}{2 m \hbar}}
(e^{i p'x/\hbar}-e^{-i p'x/\hbar})
\nonumber \\
& \times &\!\!\Bigg\{\frac{w[u_{0}(p',\tau)]\!-\!w[-u_{0}(p',\tau)]
\!-\!w[-u_{0}(p,\tau)]\!+\!w[u_{0}(p,\tau)]}{p'+p} 
\nonumber\\
& + &\!\!\frac{w[-u_{0}(p',\tau)]\!-\!w[u_{0}(p',\tau)]\!-\!w[-u_{0}(p,\tau)]
\!+\!w[u_{0}(p,\tau)]}{p'-p} \Bigg\}.
\nonumber
\end{eqnarray}
Using the symmetry under $p'\rightarrow -p'$
the previous expression is simplified to
\begin{eqnarray}
&&\la x\vert\psi^{(-1)}_{p,\tau}(t)\ra  =  \frac{i}{4 \pi }
\int_{-\infty}^{\infty}dp' 
e^{-i\frac{p'^{2}\Delta t}{2 m \hbar}+i p'x/\hbar}
\nonumber
\\
&\times&\!\!\!\Bigg\{\frac{w[u_{0}(p',\tau)]\!-\!w[-u_{0}(p',\tau)]
\!-\!w[-u_{0}(p,\tau)]\!+\!w[u_{0}(p,\tau)]}{p'+p}
\nonumber\\
& + &\!\!\frac{w[-u_{0}(p',\tau)]\!-\!w[u_{0}(p',\tau)]\!-\!w[-u_{0}(p,\tau)]
\!+\!w[u_{0}(p,\tau)]}{p'-p} \Bigg\}, 
\nonumber
\end{eqnarray}
where the relation
\beq
\label{wz}
w(z)+w(-z)=2e^{-z^{2}}
\eeq
can be applied leading to
\begin{eqnarray}
\psi^{(-1)}_{p,\tau}(x,t)& = &
\frac{-i}{2\pi }\int_{-\infty}^{\infty}dp' 
e^{-i\frac{p'^{2}\Delta t}{2 m \hbar}+i p'x/\hbar}
\nonumber
\\
&\times&\bigg\{
e^{-i\frac{p^{2}\tau}{2 m \hbar}}-w[u_{0}(p,\tau)]
-e^{-i\frac{p'^{2}\tau}{2 m \hbar}}\bigg\}
\nonumber\\
& \times & \left(\frac{1}{p'+p}+\frac{1}{p'-p}\right) + \mathcal{Y},
\end{eqnarray}
with
\beqa
\mathcal{Y}&=&-\frac{i}{2\pi }\int_{-\infty}^{\infty}dp' 
e^{-i\frac{p'^{2}\Delta t}{2 m \hbar}+i p'x/\hbar}
\nonumber
\\
&\times&\left\{\frac{w[u_{0}(p',\tau)]}{p'-p}
+\frac{w[-u_{0}(p',\tau)]}{p'+p}\right\}. 
\nonumber
\eeqa
The resulting integrals, but for the term $\mathcal{Y}$,
are of the form solved in Appendix A. 
The final result 
is Eq. (\ref{sinecase}).
\section{ 
cosine initial conditions}
The solution to the Moshisnky shutter with
cosine initial  conditions, ${\mathcal{R}}=1$,  
$$
\psi_p^{(1)}(x,t)=\frac{e^{\frac{i m x^{2}}{2 t \hbar}}}{2 }\,
\{w[-u(p,t)]+w[-u(-p,t)]\}, 
$$
%
%
is immediately computed from Eq. (\ref{mos}). 
The same result is obtained using the source boundary conditions
of Eq. (\ref{sour}) 
%
and following the steps leading to Eq. (\ref{sou}).   

In fact the same agreement can be extended to finite opening times $\tau$. 
The procedure to obtain $\psi_{p_0,\tau}^{(1)}(x,t)$ is
very much the same as for the $\emph{sine}$ case. 
Using 
\begin{displaymath}
\psi^{(1)}_{p,\tau}(x,t)=\psi^{(0)}_{p,\tau}(x,t)+
\psi^{(0)}_{-p,\tau}(x,t),
\end{displaymath}
the expressions for the $\psi^{(0)}_{\pm p,\tau}(t)$
in Eq. (\ref{psi0}), the identity in Eq. (\ref{wz}),  
%
%
%
%
and $t=\tau+\Delta t$, one finds 
\begin{eqnarray}
\psi^{(1)}_{p,\tau}(x,t)& = & \frac{i}{4\pi }\int_{-\infty}^{\infty}
dp'\, \left(e^{-i\frac{p'^{2}t}{2 m \hbar}}-e^{-i\frac{p^{2}\tau}{2 m \hbar}}
e^{-i\frac{p'^{2}\Delta t}{2 m \hbar}}\right)
\nonumber
\\
&\times&\left({e^{i p'x/\hbar}-e^{-i p'x/\hbar}}\right)
\left(\frac{1}{p'+p}+\frac{1}{p'-p}\right).
\nonumber
\end{eqnarray}
Due to the symmetry under $p'\rightarrow-p'$ the term with 
$-e^{-i p'x/\hbar}$ is equal to the one with $e^{i p'x/\hbar}$
and the resulting integral can be carried out by completing 
the square and deforming the contour of integration  in the complex 
$p$-plane as described in Appendix A, 
so new Faddeyeva functions can be identified.
This leads to the same expression found in Eq. (\ref{eq:cos}).

\end{appendix}

\end{document}